\documentclass[prl,twocolumn,groupedaddress,showpacs]{revtex4}
\usepackage{graphicx}
\begin{document}
\bibliographystyle{apsrev}


\title{Population inversion through charge measurement
using a superconducting single-electron transistor biased
in the subgap regime}
\date{\today}



\author{G\"oran Johansson}
\email[]{goran@tfp.physik.uni-karlsruhe.de}
\affiliation{Institut f\"ur Theoretische Festk\"orperphysik, 
Universit\"at Karlsruhe, D-761 28 Karlsruhe, Germany}


\date{\today}

\begin{abstract}
We show how population inversion (PI) occurs in a
two-level system (TLS) while measuring its charge using a capacitively coupled 
superconducting single-electron transistor (SSET), biased in the 
subgap regime, where the current through the SSET is carried by 
different cycles involving tunneling of both Cooper pairs and 
quasiparticles. The PI is directly associated with the resonant
nature of the Cooper-pair tunneling. We also show how the SSET may
strongly relax the TLS, although there is negligible current flowing
through the SSET, i.e. it is turned off.
The calculation of the quantum back-action noise is based on 
a real-time Keldysh approach.
We specifically discuss the case of a Cooper Pair Box qubit
with the SSET capacitively coupled as read-out device.
\end{abstract}

\maketitle

The single-electron transistor (SET) is known as an extremely
sensitive electrometer \cite{Fulton,Likharev} based on the Coulomb
blockade\cite{Averin&Likharev,Grabert}. The SET consists
of a small capacitance island, isolated from the leads
by tunnel junctions. The measurement is performed by measuring 
the current flowing through the island, which depends strongly on 
the charge induced through the capacitive coupling 
to the charge to be measured.

With the invention of the RF-SET\cite{SchoelkopfSc},
the SET was also made fast, and operating frequencies 
above 100 MHz could be reached.
Therefore the SET can be used in applications 
ranging from very sensitive charge meters and current standards\cite{Grabert} 
in which electrons are counted or pumped one by one, to read-out of
quantum bits\cite{AassimePRL,Lea,KanePRB,MakhlinRMP}
or to work as photon detectors\cite{SchoelkopfIEEE}.

For these sensitive applications the back-action of the
SET during read-out, arising from the voltage fluctuations
on the SET island, is important since it may corrupt the
measurement\cite{NazarovJLTP,JohanssonPRL}.

The SET may also be used in the superconducting state (SSET)
and indeed such a SSET shows the state-of-the-art
sensitivity of $3.2\cdot10^{-6}\mathrm{e}/\sqrt{\mathrm{Hz}}$
\cite{AassimeAPL}. This SSET was biased around the gap edge, i.e.
the threshold voltage for sequential quasiparticle tunneling\cite{KorotkovAPL},
where the back-action may be analyzed in the same way
as for sequential electron tunneling in the normal state
\cite{NazarovJLTP,JohanssonPRL,KackPRB}.

There have also been suggestions for using the SSET biased
in the subgap regime\cite{Lehnert,GunnarssonLT}, where the current
is carried by different sequences of resonant Cooper-pair
tunneling and quasiparticle tunneling, so called 
Josephson-Quasiparticle (JQP) cycles\cite{Fulton,AMvdB}.

\begin{figure}
\includegraphics[width=7cm]{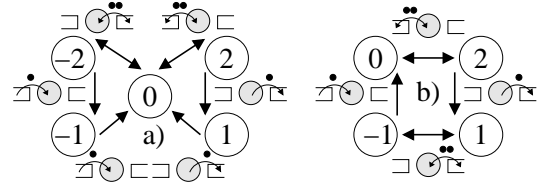}%
\caption{
a) Two simple JQP processes in parallel.
The Cooper-pair tunneling in the left/right cycle 
is resonant along the line $eV_{ds}=4(1\pm n_x)E_C$.
b) The Double JQP cycle, resonant around $eV_{ds}=2E_C$ and $n_x=0.5$.
}
\label{TJQP+EYE}
\end{figure}

The current noise, determining the shot-noise limited
sensitivity was recently analyzed for the simplest JQP cycle\cite{ChoiPRL}.
This process consists of a Cooper-pair tunneling across one junction
followed by two quasiparticles tunneling across the opposite junction.
Clerk et al\cite{ClerkPRL} analyzed a more complicated
process with two Cooper pairs involved, called
the Double JQP process (DJQP), which corresponds to the 
so called eye-feature in the SSET current-voltage characteristics,
see Fig.~\ref{TJQP+EYE}b. 
The resonant Cooper-pair tunneling was shown to improve sensitivity.

For the DJQP it was shown that also the back-action 
changes qualitatively and may even induce population inversion
(PI) in the measured system\cite{ClerkPRL}. This shows
that the back-action of the SSET indeed is very nontrivial,
e.g. simply assigning a noise-temperature to this device
would demand negative temperatures.

In this paper we describe the back-action in the whole subgap
regime. We specifically map out where PI is possible, which also includes
the simplest JQP process. Furthermore, we show where the
relaxation of the measured system may be strong, even though there
is negligible current flowing through the SSET, i.e. it is turned off.
We choose the Single Cooper Pair box (SCB)\cite{BouchiatSCB} as 
our specific two-level system (TLS).

\begin{figure}
\includegraphics[width=7cm]{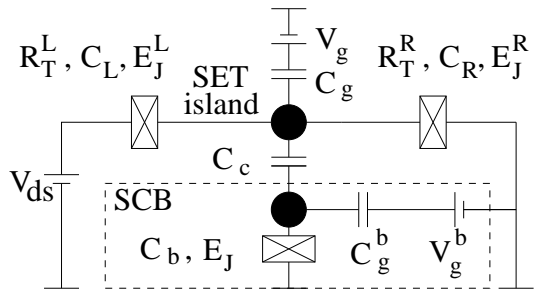}%
\caption{Schematic figure of the superconducting single-electron transistor 
capacitively coupled to a two-level system, in this case a 
Single Cooper Pair box (SCB).}
\label{qnoise_SET_fig}
\end{figure}

\paragraph{The Model}
Consider a small superconducting SET island coupled via low transparency
tunnel barriers to two external superconducting leads, and coupled
capacitively to 
a control gate and to the SCB, see Fig.\ref{qnoise_SET_fig}.
The voltage noise on the SET island is calculated neglecting
the coupling to the SCB. This approach is appropriate in the
considered limit of weak SET-box coupling ($C_c \ll C_L \sim C_R \sim C_{b}$).

We follow the outline of Ref.~\cite{AMvdB} and model
the SSET by the Hamiltonian
\begin{equation}
H=H_L+H_R+H_I+H_C+K+H_J+H_T=H_0+H_T,
\end{equation}
where $H_r=\sum_{kn}\epsilon^r_{kn}a^\dagger_{krn}a_{krn}$ and
$H_I=\sum_{ln}\epsilon_{ln} c^\dagger_{ln} c_{ln}$
describe the noninteracting quasiparticles in the $r\in\{L,R\}$
leads and on the island.
Introducing the operators $\hat{n}_{L/R}$ for the
number of charges passed from left to right in 
the $L/R$ junction we write the Coulomb term
$H_C=E_C(\hat{N}-{n_x})^2$ where $E_C=e^2/2C_\Sigma$ is the 
charging energy and $\hat{N}=\hat{n}_L-\hat{n}_R$ is the operator
for the excess number of charges on the island.
%
%
$C_\Sigma\approx C_L+C_R$ is the island capacitance 
neglecting the small gate and coupling capacitances, and
$n_x$ is the fractional number of electrons induced by the
gate voltage $C_g V_g/e$.
The work done by the biasing circuit is
$K=eV_{ds}(C_R \hat{n}_L +C_L \hat{n}_R)/C_\Sigma$
and the Josephson coupling is $H_J=-\sum_r E^r_J \cos 2\Phi_r$,
where $e^{i \Phi_r}$ increases $n_r$ by one.
Then we work in the eigenbasis of $H_0$, and treat the quasiparticle
tunneling $H_T$ as the perturbation. We use a real-time diagrammatic
Keldysh technique\cite{SchoellerSchoen,JohanssonPRL} to
calculate the DC current and the spectral density of the SET island 
voltage fluctuations,
\begin{equation}
\label{noisedef}
S_V(\omega) = \frac{e^2}{C_\Sigma^2} \int_{-\infty}^\infty d\tau
e^{+i\omega\tau} Tr\{\rho_{st}(t_0) \hat{N}(\tau) \hat{N}(0)\},
\end{equation}
which determine the back-action.
We consider lowest non-vanishing order in $H_T$, i.e. we neglect
co-tunneling and higher order effects. Due to the low junction
transparency we also neglect logarithmic renormalization effects
\cite{SchoellerSchoen}. In this approximation the rates for the
quasiparticle processes involved are given by the
usual current-voltage characteristics for tunneling between two
superconductors\cite{Tinkham}, where the effective voltage
is given by the energy gain in the process.

Assuming that the SSET is the strongest noise source coupled to
the two-level system (TLS) the rate for relaxing/exciting the TLS
is given by\cite{MakhlinRMP,DevoretNature,AguadoPRL}:
\begin{equation}
\Gamma_{\downarrow/\uparrow}= \alpha \frac{e^2}{\hbar^2} S_V(\pm\Delta E/\hbar)
\end{equation}
where $\alpha$ is a dimensionless coupling constant and 
$\Delta E$ is the energy
splitting between the two states in the TLS. For the SCB
$\alpha=\frac{C_c^2}{C_b^2}\frac{E_J^2}{\Delta E^2}$.
To describe population inversion we calculate the polarization
of the TLS in the steady state, i.e. 
$\Delta P=P^{TLS}_1-P^{TLS}_0=
(\Gamma_{\uparrow}-\Gamma_{\downarrow})/(\Gamma_{\uparrow}+\Gamma_{\downarrow})$.
The total back-action induced mixing rate 
$\Gamma_{\uparrow}+\Gamma_{\downarrow}$
is proportional to the symmetrized noise 
$S_V^{sym}(\Delta E/\hbar)=S_V(\Delta E/\hbar)+S_V(-\Delta E/\hbar)$.
If this rate is small, then other noise sources may determine 
the steady-state of the two-level system. To see where the back-action 
is strong we also study $S_V^{sym}(\Delta E/\hbar)$.

\begin{figure}
\includegraphics[height=6.5cm]{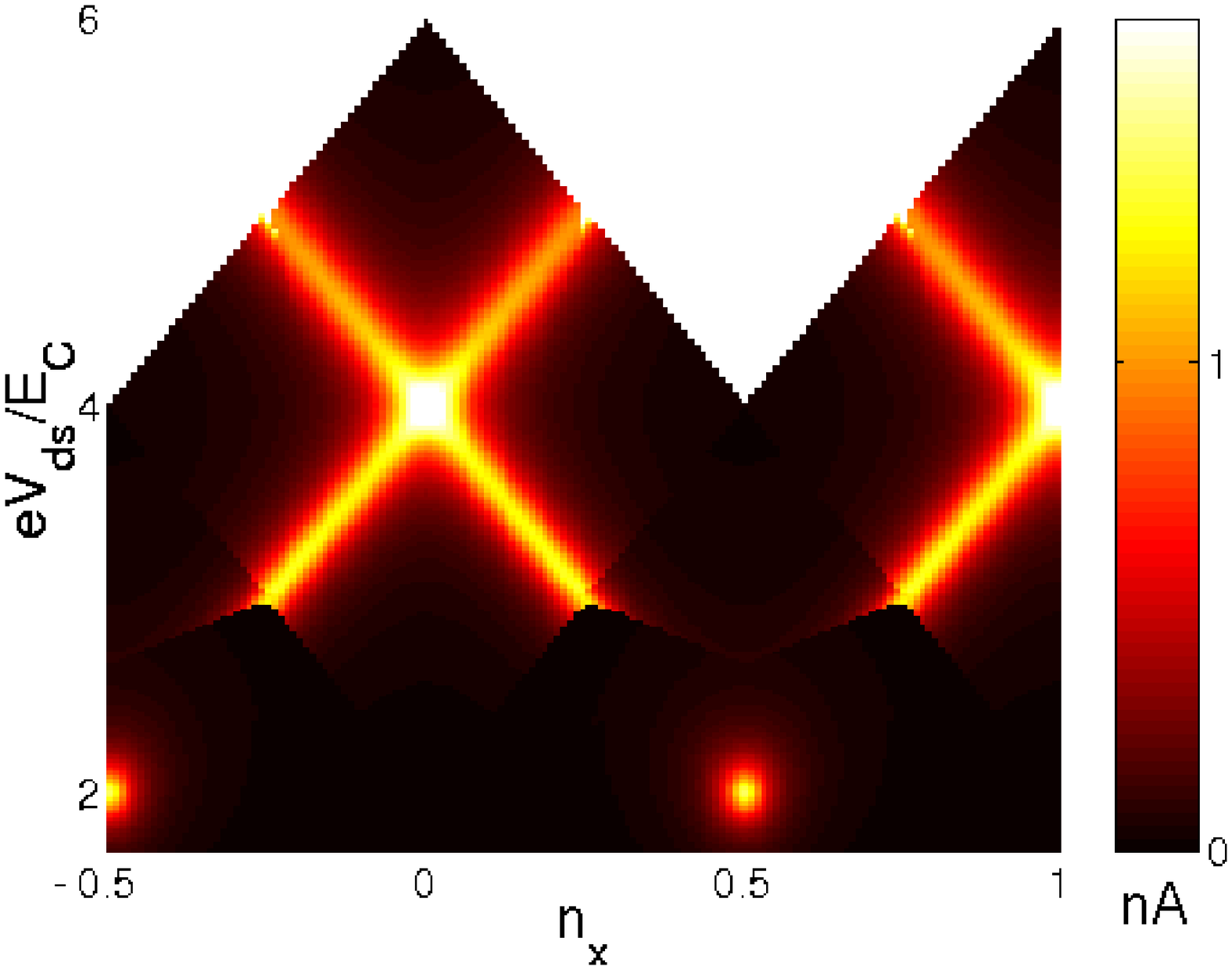}\\
\includegraphics[height=6.5cm]{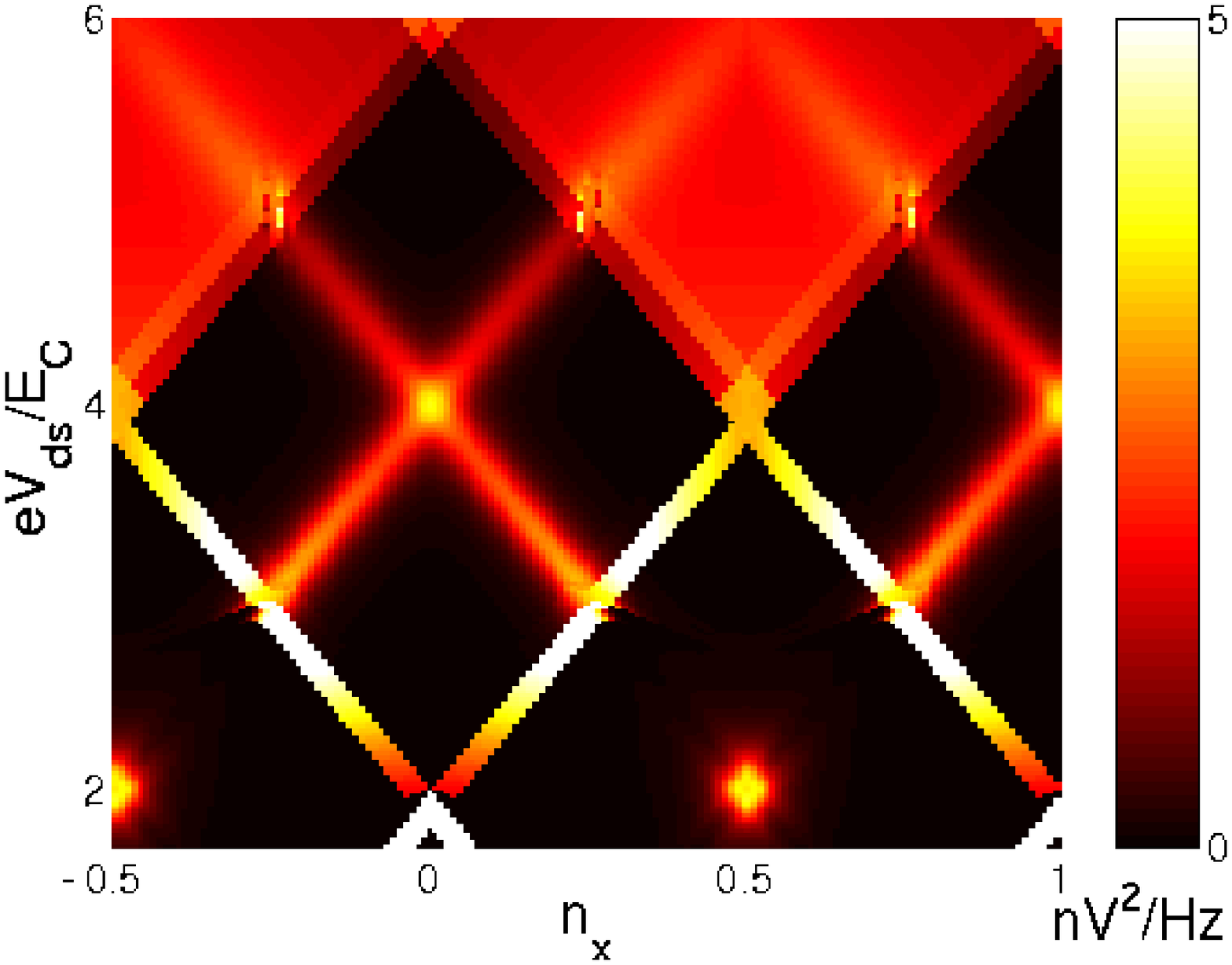}\\
\includegraphics[height=6.5cm]{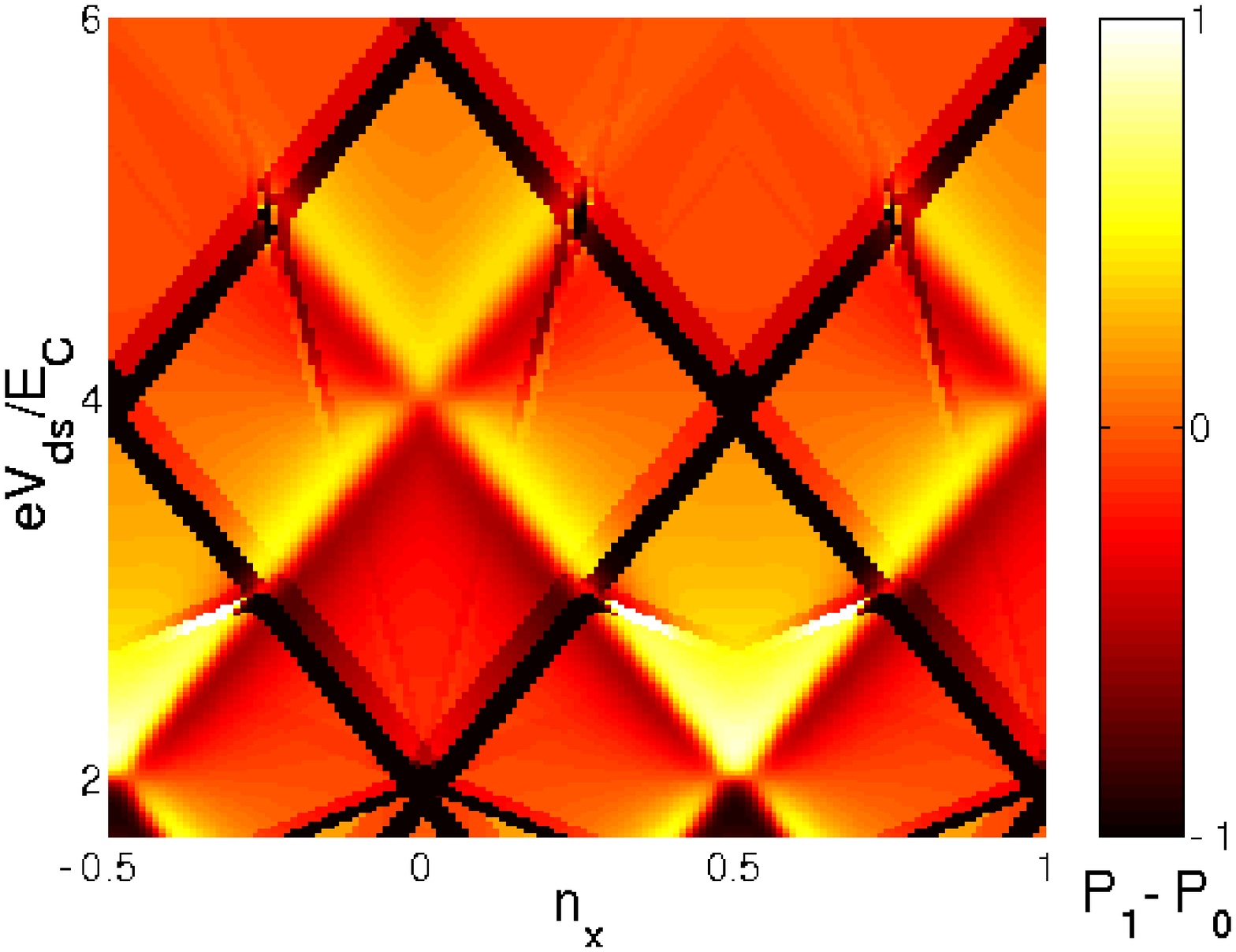}%
\caption{Numerical results for a symmetric SSET with 
$E_C=\Delta_S\approx 25\mathrm{meV}$,
$R_T=R_L\approx 50\mathrm{k}\Omega$ and $E_J^L=E_J^R\approx16\mathrm{\mu eV}$.
The level splitting in the two-level system is $\Delta E=0.1\Delta_S$.
Top panel - DC current through the SSET.
Middle panel - Total noise $S_V^{sym}(\Delta E/\hbar)$.
Lower panel - TLS population polarization $\Delta P=P_1-P_0$. 
}
\label{NumRes}
\end{figure}

\paragraph{Results}
To be specific we now consider a symmetric SSET with the same parameters
as in reference \cite{ClerkPRL}, i.e. with charging energy
equal to half of the superconducting energy gap, 
$E_C=\Delta_S\approx 25\mathrm{meV}$, $C_L=C_R$,
and $R_T=R_L\approx 50\mathrm{k}\Omega$, using the Ambegaokar-Baratoff value
for the Josephson energy, $E_J^L=E_J^R\approx16\mathrm{\mu eV}$.
We choose a level splitting of the TLS of $\Delta E=0.1 \Delta_S$.
To describe the main features in the current through the SSET and the 
noise in this regime it is enough to take the five lowest-energy
charge states into account. Numerically we check that including
7 states does not change the result.
In Fig.~\ref{NumRes} we show the numerical results for
the DC current through the SSET, the symmetrized 
noise $S_V^{sym}(\Delta E/\hbar)$ and
the polarization of the TLS $\Delta P$, as a function of
SSET gate voltage and driving bias. 

\paragraph{Analytical Results}
At specific bias points in the subgap regime,
where specific JQP processes are known to dominate, one
may calculate $S_V(\omega)$ analytically. The main
approximation is to consider all involved quasiparticle
transition rates equal and frequency independent.
For frequencies not too high compared to the superconducting
energy gap ($\hbar\omega < \Delta_S$) this is a reasonable
approximation. Keeping track of the density matrix of up to 
5 charge states, and the coherence in two resonant pairs,
leads to inverting matrices of dimension up to $9\times9$. 
Using Mathematica we obtain analytical
results, which are presented below at appropriate places.
For the DJQP cycle, in the proper limit, this procedure reproduces 
the formulas for $S_V(\omega)$ derived by Clerk et al\cite{ClerkPRL}.

\paragraph{Simple JQP resonances}
The Coulomb energy of the charge state with $N$ extra electrons on
the island is $E_N=E_C(N-n_x)^2$. 
Along the lines $eV_{ds}=E_{\pm2}-E_0=4(1\mp n_x)E_C$ in Fig.~\ref{NumRes} 
the charge states $N=0$ and $N=\pm2$ have equal energy, including
the work $2(eV_{ds}/2)$ done by the bias voltage across the L/R junction.
(Since $C_L=C_R$ the voltage divides equally over the junctions.)
Due to the Josephson coupling Cooper pairs tunnel resonantly\cite{AverinJETP}
back and forth across the L/R junction. 
For $eV_{ds}>3E_C$ the resonantly tunneled Cooper pair may decay
into two quasiparticles tunneling across the opposite (R/L) junction, thus
completing the simplest JQP cycle\cite{Fulton}, which effectively 
transports two electrons through the SSET, see Fig.~\ref{TJQP+EYE}a.  
Approximating the two involved quasiparticle rates with a
single frequency-independent rate $\Gamma/\hbar$ and
defining the energy gain for the Cooper-pair tunneling as
$\delta_{\pm}=eV_{ds}-4(1\mp n_x)E_C$,
the population polarization close to a single resonance, 
but not close to both is
\begin{equation}
\Delta P=\frac{8\delta_{\pm}\Delta E}{\Gamma^2+4(\delta_{\pm}^2+\Delta E^2)},
\label{DPsingleJQP}
\end{equation}
with maximum $\max\{\Delta P^{JQP}\}=2\Delta E/\sqrt{\Gamma^2+4\Delta E^2}$
located at $\delta_{\pm}=\sqrt{(\Gamma/2)^2+\Delta E^2}$.
We see that for positive $\delta_{\pm}$ the population is inverted,
and that the maximum inversion grows with increased TLS
level splitting $\Delta E$.

\begin{figure}
\includegraphics[width=7cm]{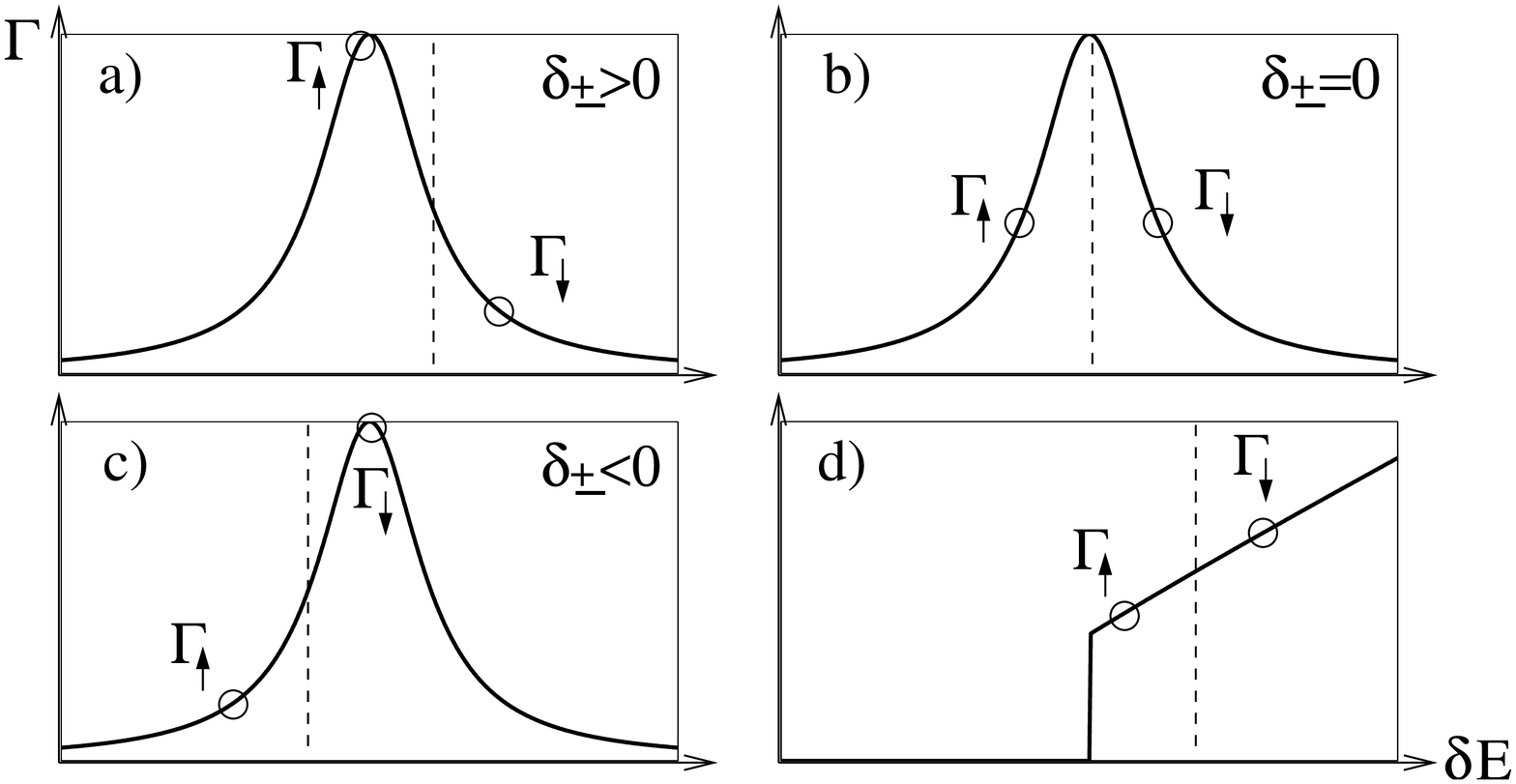}%
\caption{Schematic figure of effective tunneling rates
for Cooper pairs (a-c) and quasiparticles (d) in the SSET, 
as function of the energy gain $\delta E$ in the process.
The vertical dashed lines denote the energy gain without transitions
in the TLS, which in (a-c) is $\delta_{\pm}$. 
$\Gamma_{\uparrow/\downarrow}$ indicate tunneling rates
with simultaneous excitation/relaxation of the TLS,
where the energy gain is shifted by $\Delta E$.
}
\label{RatesFig}
\end{figure}

\paragraph{Qualitative explanation of the PI}
Without interaction with the TLS the effective Cooper-pair
tunneling rate has a maximum at zero energy gain 
$(\delta_{\pm}=0)$\cite{AverinJETP,ClerkPRL}.
In a JQP cycle which excites the TLS the effective energy
gain is lowered to $\delta_{\pm}-\Delta E$, see Fig.~\ref{RatesFig}. 
For positive $\delta_{\pm}$ this is closer to resonance (zero) than
$\delta_{\pm}+\Delta E$, which is the gain for the
relaxing cycle. Thus the exciting cycle runs faster than the
relaxing one, causing a PI of the TLS.
As for the quasiparticle tunneling rates they monotonously 
increase with increasing energy gain (see Fig.~\ref{RatesFig}d).
Thus PI will never occur above the threshold voltage, 
where the current is carried only by quasiparticles.
There the relaxing (pure quasiparticle) cycle will always
run faster than the exciting one.

\paragraph{Two JQP cycles in parallel}
Close to $eV_{ds}=4E_C$ and $n_x=0$, Cooper-pair tunneling
across both the left and right junctions are resonant simultaneously.
Here the two JQP cycles run in parallel, see Fig.~\ref{TJQP+EYE}a, 
sharing the charge state $N=0$. From the numerical results we
deduce that the maximum population inversion is not larger
at this crossing compared to the single resonances. 
Again approximating all involved quasiparticle rates with a 
single $\Gamma$, and now also taking the relevant limit $E_J \ll \Gamma$,
we find along the vertical symmetry line 
$\delta_+=\delta_-=\delta=eV_{ds}-4E_C$ and
$n_x=0$, that the population is indeed given by 
Eq.~(\ref{DPsingleJQP}). Along the horizontal line $eV_{ds}=4E_C$
we have $\delta_+=-\delta_-$ giving $\Delta P=0$.

\paragraph{The Double JQP cycle}
Around $eV_{ds}=2E_C$ and $n_x=0.5$, Cooper-pair tunneling between
the charge states $N=-1$ and $N=1$ across the right junction
and between $N=0$ and $N=2$ across the left junction are both
resonant processes. This is where the  
Double JQP process\cite{ClerkPRL} occurs, transporting 3e in each
cycle, see Fig~\ref{TJQP+EYE}b.
The population polarization here is remarkably similar to
the one around the simple JQP crossing, except that here
it has a real maximum along the symmetry line $n_x=0.5$.
For $E_J\ll\Gamma$ the maximum is again located at
$\delta=\sqrt{(\Gamma/2)^2+\Delta E^2}$, where
now $\delta=eV-2E_C$, and the value is
\begin{equation}
\Delta P^{DJQP}_{max}=
\frac{4\Delta E\sqrt{\Gamma^2+4\Delta E^2}}{\Gamma^2+8\Delta E^2}.
\end{equation}

\paragraph{Strong Relaxation - No Current}
For the SSET back-action to be important compared to other noise
sources the total noise $S_V^{sym}(\Delta E/\hbar)$ should be
large. In the middle panel of Fig.~\ref{NumRes} we see that 
the noise is large where there is substantial current, 
with one exception:
The noise is also large below the 
lines $eV_{ds}/2=E_{\pm1}-E_0=(1\pm 2 n_x)E_C$,
where the width of the region is set by the 
TLS level splitting $\Delta E$. Here a strong relaxation
of the TLS occurs, although there is negligible current.
In this bias regime the Cooper-pair tunneling is off resonance,
but it still slowly takes the SSET from the lowest energy charge state
$N=0$ to higher energy states.
Above these lines the SSET relaxes through quasiparticle tunneling.
Thus the SSET mainly occupies its lowest energy charge state $N=0$.
Below these lines this relaxation of the SSET is not energetically
allowed, and the SSET ends up with a population of the excited
states $N=\pm1$ of order unity.
In the region with strong noise the quasiparticle
transition $N=\pm1$ to $N=0$ is possible through
absorption of the TLS energy, thus causing
strong TLS relaxation although there is negligible 
current flowing through the SSET.
In short, the slow off-resonance Cooper-pair tunneling pumps the
SSET to an excited state which may only relax through a
simultaneous relaxation of the TLS.

\begin{figure}
\includegraphics[width=7cm]{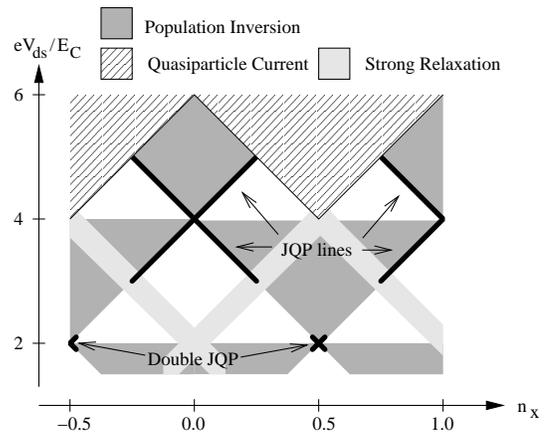}%
\caption{The dark grey indicates for which SSET bias
PI of the TLS occurs. Thick black lines indicate substantial 
subgap current through the SSET.
The light shaded areas indicate where strong TLS relaxation
occurs, although there is negligible current running 
through the SSET.
}
\label{PI_Diamonds}
\end{figure}

\paragraph{Conclusion} 
Using a real-time diagrammatic Keldysh approach we have
calculated the back-action noise of a superconducting
single-electron transistor, biased in the subgap regime.
We find that population inversion of the measured two-level
system is possible when the JQP cycle which
gives energy to the TLS is closer to resonance than the
JQP cycle which takes energy from the TLS. We also find
regions where the relaxation of the TLS is strong, although
there is negligible current flowing through the SSET,
see Fig.~\ref{PI_Diamonds}.
%
\begin{acknowledgments}
The author acknowledges fruitful discussions with
Yuriy Makhlin, Alexander Shnirman, Per Delsing, Tim Duty, and Andreas K\"ack. 
This work was supported by
the Humboldt Foundation, the BMBF and the ZIP programme of
the German government, and by the European Union under the SQUBIT project.
\end{acknowledgments}

\end{document}